\documentclass[sigconf]{acmart}
\usepackage{sistyle}
\usepackage{pgfplots}
\usetikzlibrary{patterns}
\usepackage{filecontents}
\SIthousandsep{,}
\usepackage{enumitem}

\newcommand{\acmref}{\fontsize{8pt}{8pt} \selectfont}

\makeatletter
\g@addto@macro\normalsize{%
  \abovedisplayskip 0pt plus2pt 
  \belowdisplayskip
  \abovedisplayskip
  \abovedisplayshortskip  0pt plus2pt%
  \belowdisplayshortskip  0pt plus0pt
}
\makeatother
\definecolor{green1}{RGB}{204,203,102}
\definecolor{blue1}{RGB}{102,204,182}
\definecolor{red1}{RGB}{102,163,204}
\definecolor{purple1}{RGB}{102,33,102}
\definecolor{green2}{RGB}{255,153,102}
\definecolor{blue2}{RGB}{51,204,204}
\definecolor{red2}{RGB}{143,153,204}

\definecolor{tuatara}{RGB}{67, 67, 67}
\definecolor{aluminum}{RGB}{153,153,153}
\definecolor{silver}{RGB}{191,191,191}
\definecolor{platinum}{RGB}{228,227,228}
\definecolor{mercury}{RGB}{240,240,240}
\definecolor{gallery}{RGB}{250,250,250}
\definecolor{free_speech_aquamarine}{RGB}{0, 156, 114}
\definecolor{sun_shade}{RGB}{255, 144, 68}
\definecolor{fern}{RGB}{101,197,117}
\definecolor{french_blue}{RGB}{0, 112, 182}
\definecolor{sushi}{RGB}{117, 168, 47}
\definecolor{shakespeare}{RGB}{35, 184, 223}
\definecolor{egg_shell}{RGB}{238, 234, 215}
\definecolor{carnation}{RGB}{245, 80, 86}
\definecolor{flamingo}{RGB}{237, 88, 85}
\definecolor{jet_stream}{RGB}{188, 214, 210}
\definecolor{jelly_bean}{RGB}{45, 126, 150}
\definecolor{tree_poppy}{RGB}{246, 154, 27}

\makeatletter
\g@addto@macro\normalsize{%
  \abovedisplayskip 2pt plus1pt 
  \belowdisplayskip 2pt plus1pt
  \abovedisplayshortskip  3pt plus1pt%
  \belowdisplayshortskip  3pt plus1pt
}
\makeatother

\renewcommand{\arraystretch}{1.2}
\begin{document}
\fancyhead{}

\title{Match$^2$: A Matching over Matching Model\\for Similar Question Identification}

\author{Zizhen Wang$^{\ast,\dagger}$, Yixing Fan$^\dagger$, Jiafeng Guo$^{\ast,\dagger}$, Liu Yang$^\ddagger$}
\email{{wangzizhen, fanyixing, guojiafeng, zhangruqing, lanyanyan, cxq}@ict.ac.cn}
\author{Ruqing Zhang$^\dagger$, Yanyan Lan$^{\ast,\dagger}$, Xueqi Cheng$^{\ast,\dagger}$, Hui Jiang$^\S$, Xiaozhao Wang$^\S$}
\email{lyang@cs.umass.edu, {huijiang.jh, orlando}@alibaba-inc.com}
\affiliation{
  \institution{
    $^\ast$University of Chinese Academy of Sciences, Beijing, China\\
    $^{\dagger}$CAS Key Lab of Network Data Science and Technology, Institute of Computing Technology, \\ Chinese Academy of Sciences, Beijing, China\\
    $^\ddagger$University of Massachusetts Amherst, Massachusetts, United States\\
    $^\S$Alibaba Group, Beijing, China
  }
}


\begin{abstract}
Community Question Answering (CQA) has become a primary means for people to acquire knowledge, where people are free to ask questions or submit answers. To enhance the efficiency of the service, similar question identification becomes a core task in CQA which aims to find a similar question from the archived repository whenever a new question is asked. However, it has long been a challenge to properly measure the similarity between two questions due to the inherent variation of natural language, i.e., there could be different ways to ask a same question or different questions sharing similar expressions. To alleviate this problem, it is natural to involve the existing answers for the enrichment of the archived questions. Traditional methods typically take a \textit{one-side} usage, which leverages the answer as some expanded representation of the corresponding question. Unfortunately, this may introduce unexpected noises into the similarity computation since answers are often long and diverse, leading to inferior performance. In this work, we propose a \textit{two-side} usage, which leverages the answer as a bridge of the two questions. The key idea is based on our observation that similar questions could be addressed by  similar parts of the answer while different questions may not. In other words, we can compare the matching patterns of the two questions over the same answer to measure their similarity. In this way, we propose a novel matching over matching model, namely Match$^2$, which compares the matching patterns between two question-answer pairs for similar question identification. Empirical experiments on two benchmark datasets demonstrate that our model can significantly outperform previous state-of-the-art methods on the similar question identification task.





\end{abstract}

\ccsdesc[500]{Information systems~Question answering}

\keywords{community question answering, similar question identification; matching over matching} 

\maketitle


{\acmref \textbf{ACM Reference Format:}\\Zizhen Wang, Yixing Fan, Jiafeng Guo, Liu Yang, Ruqing Zhang, Yanyan Lan, Xueqi Cheng, Hui Jiang and Xiaozhao Wang. 2020. Match2: A Matching over Matching Model for Similar Question Identification. In \textit{43rd International ACM SIGIR Conference on Research and Development in Information Retrieval (SIGIR 20), July 25 to July 30, 2020, Virtual Event, China.} ACM, New York, NY, USA, 10 pages. https://doi.org/10.1145/3397271.3401143}

\section{Introduction}

\begin{figure*}[h]
  \centering
  \includegraphics[width=0.8\linewidth]{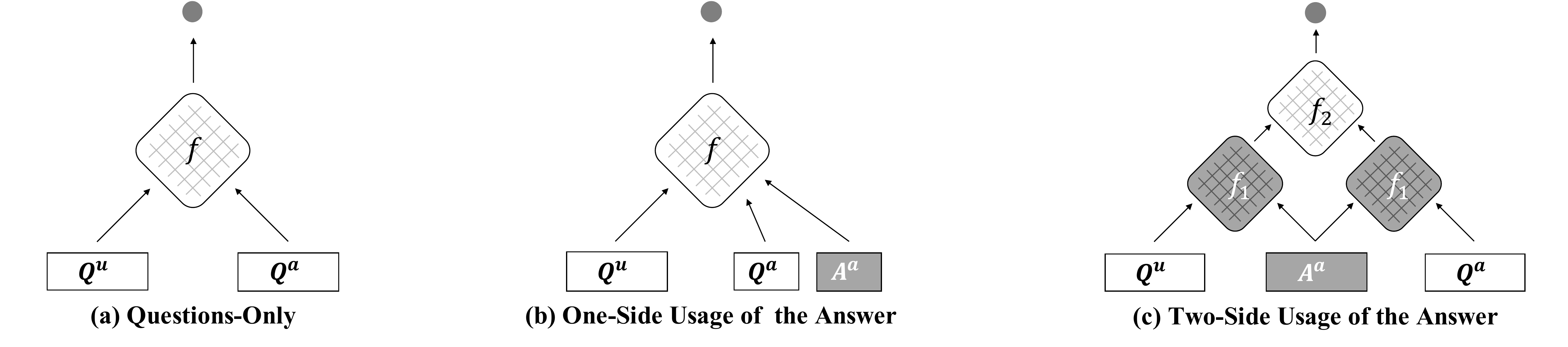}
  \caption{The architectures of similar question identification models. $f$ denotes the identification function. The first architecture only uses the questions for identification, while the last two involve the archived answer, in which (b) treats the answer as an expand to the question and (c) leverages the answer as a bridge of the two questions.}
  \label{fig:ans}
\end{figure*}

\begin{figure}[h]
  \centering
  \includegraphics[width=\linewidth]{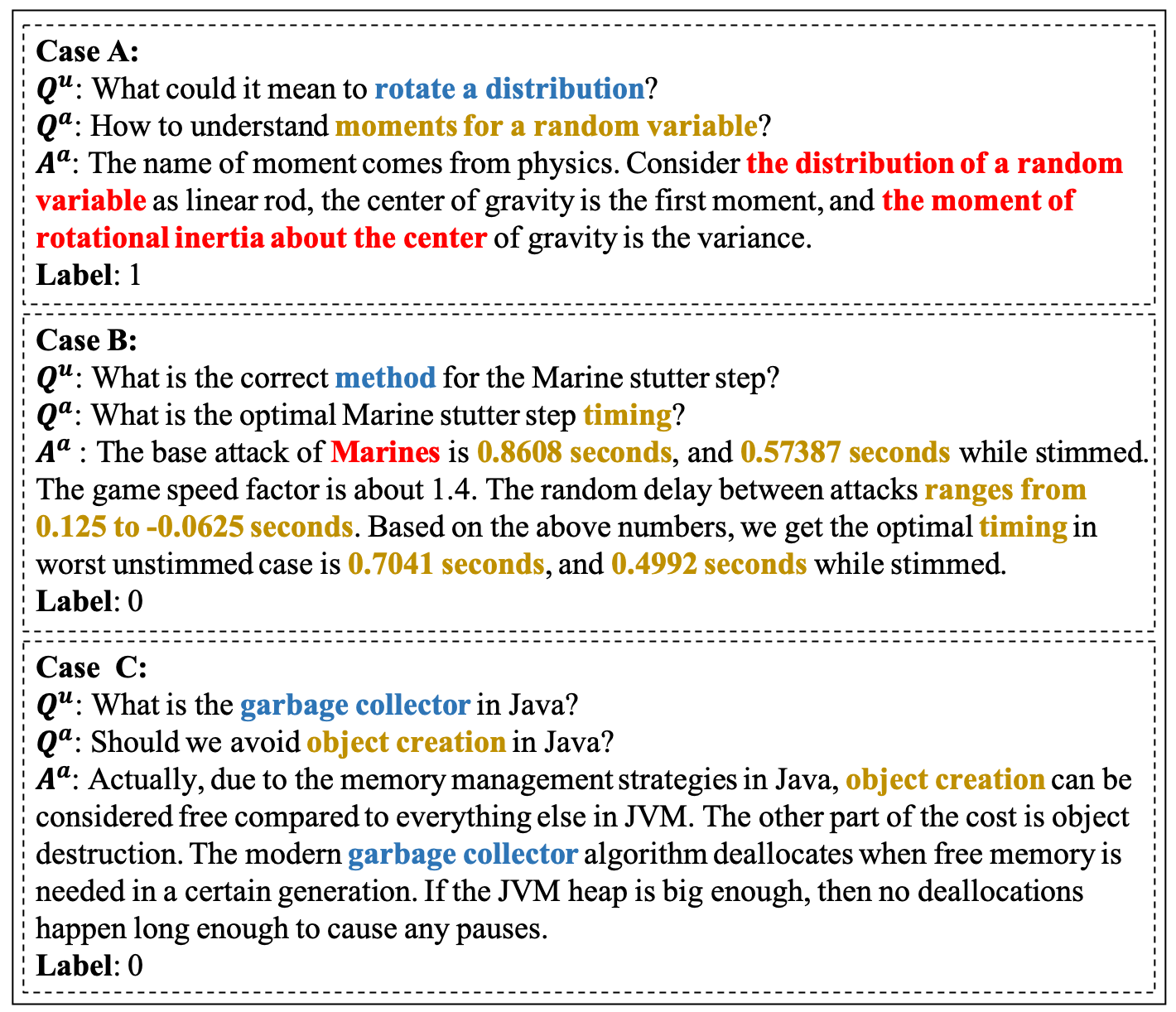}
  \caption{The cases from StackExchange. The blue and yellow parts denote the focuses of the question and the corresponding related answer parts. The red parts denote the answer text which can address both the user question and the archived question. The archived answer is helpful to identify the question similarity (Case A,B), but it may introduce unexpected noises (Case C).}
  \label{fig:intro}
\end{figure}

Community Question Answering (CQA) services, such as StackExchange\footnote{https://stackexchange.com/} and Quora\footnote{https://www.quora.com/}, have grown in popularity in recent years as a platform for people to exchange knowledge. In CQA, users can ask their questions or submit answers to questions in a collaborative fashion.
Although CQA services greatly benefit users with high-quality human-generated answers for solving their problems, the efficiency becomes a big concern as the asker need to wait until someone submits the answer to his/her question. To alleviate this problem, similar question identification becomes a core task in CQA which aims to find a similar question from the archived repository whenever a new question is proposed. In the meantime, similar question identification could also help reduce redundant questions in CQA services, saving a lot of users' efforts. 


However, it has long been a challenge to properly measure the similarity between two questions, which are usually very short in length, due to the inherent variation of natural language. On one hand, there could be different ways to express the same question, leading to the lexical gap \cite{ji2012question,qiu2015convolutional}. For example, as shown in Figure \ref{fig:intro} Case A, the user question $Q^u$ and the archived question $Q^a$ are similar and could be addressed by the archived answer $A^a$ of $Q^a$, but they have very different expressions. On the other hand, there could be different questions sharing very similar expressions, leading to false positive predictions if one cannot distinguish their subtle difference. For example, as shown in Figure \ref{fig:intro} Case B, although these two questions share many words in common, their focus is totally different (one about the ``method'' and one about the ``timing''), and thus could not be addressed by the same answer.


Similar question identification has attracted extensive studies in recent years. Some early works in this direction formulated it as a question-question matching problem, as shown in Figure \ref{fig:ans}(a). Both conventional machine learning methods \cite{cao2008recommending,yang2013cqarank,xue2008retrieval,guzman2019machine} and deep neural networks \cite{nakov2016takes,qiu2015convolutional,wan2016match,chen2016enhanced,gong2017natural,yang2019simple,devlin2018bert} have been applied to this problem. However, simply based on two questions, even most advanced neural models cannot well address the two challenges mentioned above due to the sparse information in questions. 

Since archived questions usually associated with answers, it is natural to involve the existing answers for the enrichment of the archived questions to alleviate the sparsity problem. To leverage the answers of the archived question, existing methods typically take a \textit{one-side} usage, as shown in Figure \ref{fig:ans}(b), which treats the answer as some expanded representation of the corresponding question.
For example, Ji et al. \cite{ji2012question} employed the archived answer to learn an enriched topic representation of questions for similarity computation.
Gupta et al. \cite{gupta2019faq} matched the user question to the archived question and its answer separately then aggregated them with an attention mechanism.
Unfortunately, the one-side usage may introduce unexpected noises into the similarity computation, leading to inferior performance. The reason is that answers are not equivalent representations of the corresponding questions. Answers are often long and cover diverse topics/aspects that may be beyond the scope of the corresponding question. For example, as shown in Figure \ref{fig:intro} Case C, these two questions, one about garbage collector and one about object creation, are different in semantics. However, if we simply expand the archived question $Q^a$ with its answer $A^a$ which also talks about the garbage collector, we are prone to predict that these two questions are similar which is apparently a false positive prediction.


In fact, if we look at these cases carefully, we may find the following observation: similar questions could be addressed by similar parts of the answer while different questions may not. For example, as shown in Figure \ref{fig:intro} Case A, these two questions are similar since they both can be addressed/connected by the similar parts of the archived answer $A^a$. However, in Case C, although the archived answer may be related to both questions, the related parts are quite different for the two questions. These cases show that it is not how similar the archived answer to the user question decides the question similarity. It is how the archived answer matches the two questions contributes to the similarity of the two questions. Therefore, we argue that the archived answer should not be simply viewed as some expansion of the corresponding question, but rather be viewed as a bridge of the two questions, namely a \textit{two-side} usage in this work as depicted in Figure \ref{fig:ans}(c).

Based on the above idea, we propose a novel Matching over Matching Model, namely Match$^2$ for short, which compares the matching patterns of the two questions over the same answer for similar question identification.
Specifically, Match$^2$ contains three modules, including the \textit{Representation-based Similarity Module}, the \textit{Matching Pattern-based Similarity Module} and the \textit{Aggregation module}.
The Representation-based Similarity module is similar to previous question matching methods, which generates a similarity vector between two questions simply based on their representations. 
The major enhancement is the Matching Pattern-based Similarity module. This module has a Siamese Network structure, which takes two question-answer pairs as the inputs, learns their matching patterns separately, builds a matching similarity tensor by comparing the two matching patterns, and finally produces the similarity vector between the questions by compressing the matching similarity tensor. Both the representation-based and matching pattern-based similarity vectors are aggregated in the Aggregation module to produce the final identification prediction. The Aggregation module adopts a gate mechanism which takes the representation-based similarity as the primary one and the matching pattern-based similarity as the complementary one for the final decision. A multi-task learning strategy is employed to train the Match$^2$ model. 

We evaluate the effectiveness of the proposed Match$^2$ model based on two widely used CQA benchmarks, i.e., CQADupStack \cite{hoogeveen2015cqadupstack} and QuoraQP\footnote{https://www.kaggle.com/c/quora-question-pairs}.
To incorporate the answer information in QuoraQP, we crawled archived answers of the corresponding questions from Quora and enrich the benchmark into a new answer-expanded version, namely QuoraQP-a.
The experimental results on these two benchmarks demonstrated that our method can significantly outperform those state-of-the-art methods on the similar question identification task.

The major contributions of this paper include:
\begin{enumerate}
    \item We analyze the role of the archived answer in the similar question identification task and propose a two-side usage of the answer which leverages it as a bridge of the two questions.
    \item We propose a novel matching over matching (Match$^2$) model to compare the matching patterns of the two questions over the same answer for similar question identification.
    \item We conduct extensive comparisons and analysis against the state-of-the-art similar question identification models on benchmarks to demonstrate the effectiveness of our proposed method.
\end{enumerate}

\section{Related Work}

In this section, we briefly review the most related topics to our work in CQA, i.e., question matching.
Question matching which evaluates the similarity between two questions, could be further divided into the question deduplication task and the similar question identification task with regard to different application scenarios.

\subsection{Question Deduplication}
Question deduplication aims to merge or remove the redundant questions in the archived question threads. 
Early studies mainly focused on designing effective features to measure the similarities between two questions, such as lexical features \cite{jaccard1901etude,broder1997resemblance,gusfield1997algorithms}, syntactic features \cite{moschitti2006efficient,carmel2014improving, wang2007jeopardy}, or heuristic features \cite{barron2015thread,filice2017kelp}. 
Many recent successes on this task have been achieved by advanced neural network models. 
For example, Pang et al. \cite{pang2016text} evaluated the question similarity from hierarchical levels. 
Wan et al. \cite{wan2016match} modeled the recursive structure between question pairs with spatial RNN. 
Tay et al. \cite{tay2018co} proposed a CSRAN model to learn fine-grained question matching details. 
Yang et al. \cite{yang2019simple} built RE2 model with stacked alignment layers to keep the model fast while still yielding strong performance, and Devlin et al. \cite{devlin2018bert} pre-trained a stacked transformer network which can be used for question deduplication task after fine-tuning. 

Besides, the question threads in the community include not only the question texts but also other information, e.g., topics, comments and answers, which provide other perspectives for question deduplication. 
Zhang et al. \cite{zhang2014question} proposed a topic model approach to take answer quality into account.
Wu et al. \cite{wu2018question} proposed the QCN network to make use of the subject-body relationship of the community questions.
Filice et al. \cite{filice2017kelp} proposed a method to utilize the interconnection information between the question and its comments. 
Liang et al. \cite{liang2019adaptive} employed adaptive multi-attention mechanism to enhance questions with their corresponding answers.
Moreover, many researchers have considered the use of different kinds of external resources. 
Wu et al. \cite{wu2017ecnu} employed various types of handcraft features to measure the question semantic similarity.
Zhou et al. \cite{zhou2013improving} used the semantic relations extracted from the global knowledge of WikiPedia\footnote{https://en.wikipedia.org/wiki/Main\_Page}.

\subsection{Similar Question Identification}

Similar question identification aims to find a similar question from the archived repository for a new question issued by a user.  
It is usual to frame the similar question identification as a retrieval task where the user question is taken as a query and archived questions are ranked based on their semantic similarities to the query.
Hence, classical retrieval methods, e.g., BM25 \cite{robertson1995okapi} and LMIR \cite{zhai2004study}, have beed applied for this task.  
There are also researchers employed statistical translation \cite{jeon2005finding,xue2008retrieval,zhou2011phrase}, topic model \cite{cai-etal-2011-learning,zhou2011learning} and relation extraction methods \cite{romano2006investigating} to identify the similar questions.
Recently, deep learning methods have been widely adopted to solve it. 
For example, Qiu et al. \cite{qiu2015convolutional} employed convolutional neural network to encode questions in semantic space.
Wan et al. \cite{wan2016deep} proposed MV-LSTM to capture the contextualized local information with multiple positional question representations.
Furthermore, many works considered the use of different kinds of complementary information, such as question category \cite{cao2012approaches,zhou2013towards,duan2008searching}, Wikipedia concepts \cite{ahasanuzzaman2016mining} and corresponding answer \cite{ji2012question,gupta2019faq,sakata2019faq}.

Even some of the researchers on similar question identification have focused on ranking models, they might face the computational complexity and evaluation difficulty problem \cite{hoogeveen2018detecting}. 
To address this issue, many works model the task as a classification task, which aims to explicitly predict whether the archived question is similar with the user question or not. 
For example, Wang et al. \cite{wang2017bilateral} employed a bilateral mechanism to enhance single direction matching.
Chen et al. \cite{chen2016enhanced} proposed a sequential inference model based on chain LSTMs for the recursive matching architectures.
Gong et al. \cite{gong2017natural} used DenseNet \cite{huang2017densely} to hierarchically extract semantic features from questions interaction space. 
Hoogeveen et al. \cite{hoogeveen2018detecting} adopted meta data such as user features to identify the question relation. 
It seems that some models could not only be applied to similar question identification but also question deduplication task, but we can find the clear difference, i.e., the user question in similar question identification has few information except the text itself.

\section{Our Approach}

\begin{table}
  \caption{A summary of key notations in this work.}
  \begin{tabular}{l|p{200pt}}
    \hline
    \hline
    $\mathbf{P}^u, \mathbf{P}^a$ & The matching pattern of the user question and the archived question over the archived answer \\
    \hline
    $\mathbf{P}^s$ & The pattern similarity tensor\\
    \hline
    $\mathcal{S}, S, s$ & The pattern similarity function at tensor-, layer- and element-wise \\
    \hline
    $\mathbf{v}_q$ & The representation-based similarity vector \\
    \hline
    $\mathbf{v}_a$ & The matching pattern-based similarity vector \\
    \hline
    $\mathbf{v}$ & The question similarity vector \\
    \hline
    $r$ & The main task loss ratio \\
    \hline
    \hline
  \end{tabular}
  \label{tab:note}
\end{table}

In this section, we present the Matching over Matching (Match$^2$) model for the similar question identification task in detail.
We first give an overview of the problem formulation and model architecture, and then describe each module of our model as well as the learning procedure. 
A summary of key notations in this work is presented in Table \ref{tab:note}.

\subsection{Overview}


Formally, given a user question $Q^u$, an archived question $Q^a$, and an answer $A^a$ of the archived question, Match$^2$ aims to learn a classification model $f(\cdot)$ to predict the similarity score $y^q$ between the user question $Q^u$ and the archived question $Q^a$. 
 
\begin{figure*}[h]
  \centering
  \includegraphics[width=0.8\linewidth]{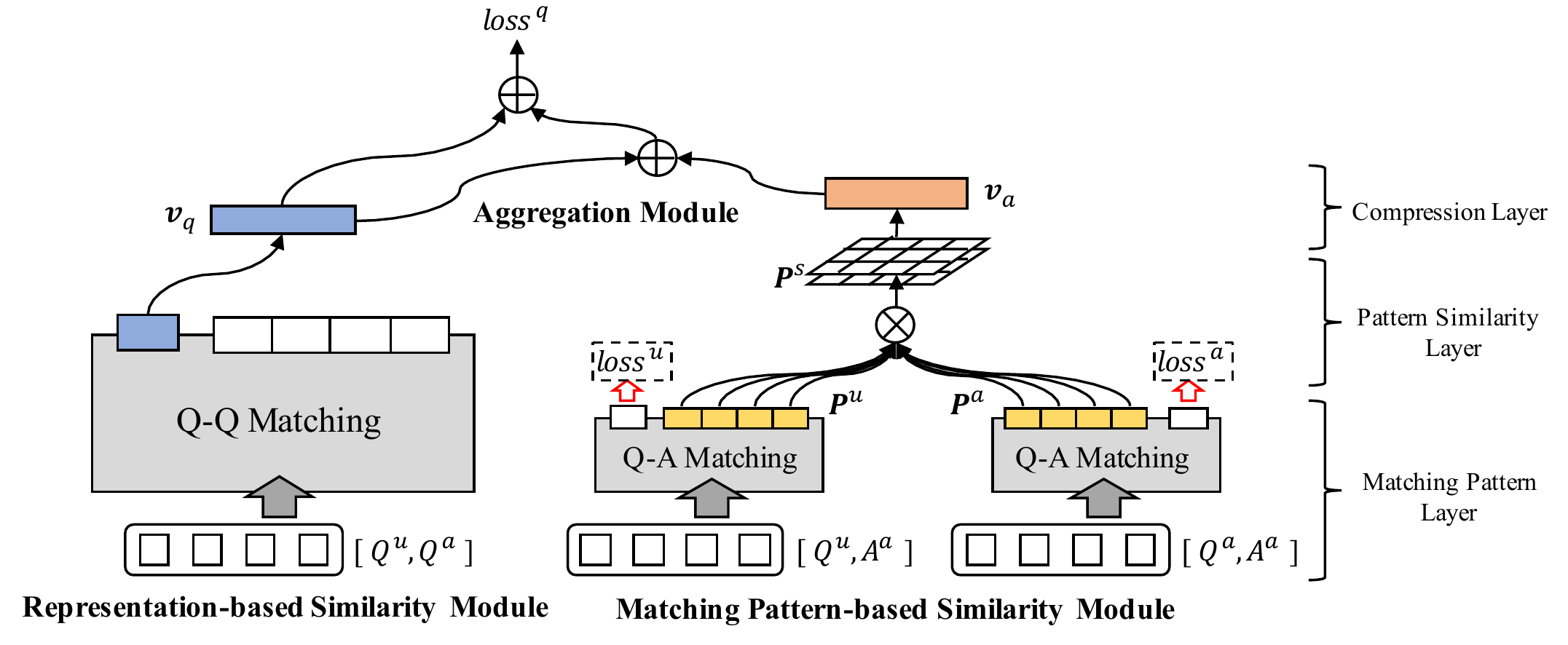}
  \caption{The architecture of the Matching over Matching Model.}
  \label{fig:arch}
\end{figure*}

Basically, our Match$^2$ model contains the following components: 
(1) Representation-based Similarity module: to produce a similarity vector between the two questions based on their representations;  
(2) Matching Pattern-based Similarity module: to compare the matching patterns of the two questions over the archived answer; 
and (3) Aggregation module: to produce the final similarity score by aggregating the representation-based and matching pattern-based module. 
The overall architecture is depicted in Figure \ref{fig:arch} and we will detail our model as follows.

\subsection{Representation-based Similarity Module} \label{sec:pooled}
Generally, the representation-based similarity module takes the two questions as inputs and predicts a similarity vector, which is similar to previous question matching methods.
In this work, we adopt the Bert \cite{devlin2018bert} to measure the question similarity due to its superiority in many natural language understanding tasks. 

Firstly, we concatenate the questions to the required format, which starts with a $[CLS]$ token for the whole sequence representation and ends with a $[SEP]$ token to denote the separator boundary of each question.
Then, we use the stacked transformer architecture to encode the formatted questions to obtain the representations.
Specifically, after being embedded, the input is processed by a multi-head attention network and a feed forward network in each transformer layer.
This stacked structure has two types of outputs,
\begin{equation}\label{eq:bert}
    \mathbf{B}^p, \mathbf{B}^s = StackedTransformer(Q^u, Q^a),
\end{equation}
where $\mathbf{B}^p$ denotes the pooled feature corresponded to $[CLS]$ and $\mathbf{B}^s$ represents the sequence features of the whole input sequence. 
We adopt $\mathbf{B}^p$ as the representation-based similarity vector $\mathbf{v}_q = \mathbf{B}^p \in R^H$, where $H$ is the hidden size of Bert.

\subsection{Matching Pattern-based Similarity Module} \label{sec:pattern}

The matching pattern-based similarity module is responsible for contrasting the matching patterns of the two questions over the same answer.  
As shown in the right part of Figure \ref{fig:arch}, this module has a Siamese Network structure, which includes three dependent layers: (1) matching pattern layer: to take two question-answer pairs as the inputs and learns their matching patterns separately; 
(2) pattern similarity layer: to build a pattern similarity tensor by contrasting the two matching patterns; 
and (3) compression layer: to produce the similarity vector between the questions by compressing the matching similarity tensor.

\subsubsection{\textbf{Matching Pattern Layer}}

We adopt Bert again to compute the matching patterns of the two questions over the same answer. 
Different from the $\mathbf{B}^p$ from Equ.\ref{eq:bert}, we use the sequence features $\mathbf{B}^s$ here and divide it into two parts to represent the question and answer respectively.

Take the matching pattern $\mathbf{P}^u$ between the user question $Q^u$ and the archived answer $A^a$ as an example. 
Firstly, the user question $Q^u$ with a sequence of $m$ tokens is represented by concatenating the question sequence features from each transformer layer, 
$$\widehat{\mathbf{Q}}^u=[\widehat{\mathbf{Q}}_1^u,\cdots,\widehat{\mathbf{Q}}_l^u,\cdots,\widehat{\mathbf{Q}}_L^u],$$
where $L$ is the number of transformer layers in Bert, $\widehat{\mathbf{Q}}_l^u \in R^{H \times m}$ is the $l$-th question sequence feature separated from $\mathbf{B}^s$. 
In the same way, the archived answer $A^a$ that has $w$ tokens is represented as $\widehat{\mathbf{A}}^a \in R^{L \times H \times w}$. 
Finally, the layer-wise matching pattern $\mathbf{P}_l^u$ between $Q^u$ and $A^a$ in the $l$-th transformer layer is computed as, 
$$\mathbf{P}_l^u=\widehat{\mathbf{Q}}^{uT}_l\widehat{\mathbf{A}}^a_l.$$
Hence, by concatenating the $L$ layer-wise matching patterns, we can obtain the final matching pattern $\mathbf{P}^u$ between the user question  $Q^u$ and the archived answer $A^a$, i.e.,
$$\mathbf{P}^u = [\mathbf{P}_1^u,\mathbf{P}_2^u,\cdots,\mathbf{P}_L^u] \in R^{L \times m \times w}$$.

The matching pattern $\mathbf{P}^a$ of the archived question answer pair can be computed in the same way as described above. 
It should be noted that the Bert architecture in this module does not share the parameters with that used in Section \ref{sec:pooled}.

\subsubsection{\textbf{Pattern Similarity Layer}} \label{sec:ps}

In this layer, we compute a pattern matching similarity tensor $\mathbf{P}^s$  given the two matching patterns $\mathbf{P}^u$ and $\mathbf{P}^a$, i.e., 
$$\mathbf{P}^s = \mathcal{S}(\mathbf{P}^u, \mathbf{P}^a) \in R^{L \times m \times n},$$
where $\mathcal{S}$ denotes the tensor-wise similarity function, and $\mathbf{P}^s_l$ denotes the layer-wise matching pattern which is defined as, $$\mathbf{P}^s_l = S(\mathbf{P}^u_l, \mathbf{P}^a_l) \in R^{m \times n}.$$
Specifically, the element-wise matching pattern similarity scalar $P^s_{l,ij}$ is computed by, 
$$P^s_{l,ij} = s(\mathbf{P}^u_{l,i}, \mathbf{P}^a_{l,j}),$$
where $\mathbf{P}^u_{l,i}$ is the matching pattern from the $i$-th token in the user question to the archived answer, as well as $\mathbf{P}^a_{l,j}$ represents that from the $j$-th archived question token.

Here, we propose five element-wise similarity functions $s(\mathbf{x},\mathbf{y})$ to compute the similarity between a question and an answer. 

\begin{itemize}[leftmargin=*]

\item \textit{Dot product} between two vectors is based on the projection of one vector onto another, which is defined as follows:
$$s_{dot}(\mathbf{x},\mathbf{y}) = <\mathbf{x}, \mathbf{y}> = \mathbf{x}^{T}\mathbf{y}.$$

\item \textit{Cosine} is a common function to model interactions. The similarity score is viewed as the angle of two vectors:
\begin{equation}\nonumber
s_{cos}(\mathbf{x},\mathbf{y}) = \frac{<\mathbf{x},\mathbf{y}>}{||\mathbf{x}|| ||\mathbf{y}||} 
\end{equation}
where $||\cdot||$ denote the $L_2$ norm of vector.

\item \textit{L1} \cite{jain2019attention} represents the similarity based on Manhattan distance between vectors as follows, 
$$s_{l1}(\mathbf{x},\mathbf{y}) = \frac{1}{1+\Sigma_{t=1} |\mathbf{x}_t - \mathbf{y}_t|}.$$

\item \textit{L2} is another widely used distance-based similarity function. Different from the \textit{L1} function, it is based on euclidean distance, namely, 
$$s_{l2}(\mathbf{x},\mathbf{y}) = \frac{1}{1+\sqrt{\Sigma_{t=1} (\mathbf{x}_t-\mathbf{y}_t)^2}}.$$

\item \textit{Jesene-Shannon} \cite{jain2019attention} firstly transforms the vector to a distribution with $softmax$ function, and then quantifies their difference by Jesene-Shannon Divergence \cite{fuglede2004jensen}, 
$$s_{jss}(\mathbf{x},\mathbf{y})=1 - JSD(softmax(\mathbf{x}), softmax(\mathbf{y})).$$

\end{itemize}

\subsubsection{\textbf{Compression Layer}} 

The compression layer aims to produce the matching pattern-based similarity vector by compressing the pattern similarity tensor $\mathbf{P}^s$ to a low dimension vector. 
We firstly use a two-layer BN-ReLU-Conv \cite{he2016deep} structure with $H$ filters to introduce contextual information, and then adopt the average global pooling method \cite{lin2013network} to obtain the final matching pattern-based similarity vector $\mathbf{v}_a \in R^H$.

\subsection{Aggregation Module}

The similarity vectors from previous two modules are combined to compute the question similarity score $y^q$ in this module. 
Given $\mathbf{v}_q$ and $\mathbf{v}_a$, we introduce a gate mechanism inspired by GRU \cite{cho2014learning}, which takes the former as the primary one and the latter as the complementary one, to obtain the final question similarity vector $\mathbf{v}$.
Specifically, it can be computed by
\begin{equation}\nonumber
    \begin{split}
        \mathbf{r} &= \sigma(\mathbf{W}_r \mathbf{v}_a+\mathbf{U}_r \mathbf{v}_q),\\
        \mathbf{z} &= \sigma(\mathbf{W}_z \mathbf{v}_a+\mathbf{U}_z \mathbf{v}_q),\\
        \widehat{\mathbf{v}} &= tanh(\mathbf{W} \mathbf{v}_a + \mathbf{U}(\mathbf{r} \otimes \mathbf{v}_q)),\\
        \mathbf{v} &= \mathbf{z} \mathbf{v}_q + (1-\mathbf{z})\widehat{\mathbf{v}},
    \end{split}
\end{equation}
where $\otimes$ is the element-wise multiplication, $\sigma$ denotes the sigmoid function, and $\mathbf{W}_r, \mathbf{W}_z, \mathbf{W}, \mathbf{U}_r, \mathbf{U}_z, \mathbf{U}$ are trainable parameters.

Based on the question similarity vector $\mathbf{v}$, we then apply a multi-layer perceptron (MLP) to obtain the question similarity score $y^q$,
\begin{equation}\label{eq:out}
  y^q = \sigma(\mathbf{W}_2 ReLU(\mathbf{W}_1 \mathbf{v}+\mathbf{b}_1) + \mathbf{b}_2),
\end{equation}
in which $\mathbf{W}_1$, $\mathbf{b}_1$, $\mathbf{W}_2$ and $\mathbf{b}_2$ are trainable parameters.

\subsection{Model Training and Inference} \label{sec:multi}

In the training phase, we employ the cross-entropy loss to learn our Match$^2$ model in an end-to-end way. 
To train the model sufficiently, we adopt a multi-task learning strategy to combine the question-question matching task and the question-answer matching task.  
The question-question matching task aims to measure the similarity between two questions as our main task, while the question-answer matching is an auxiliary task that aims to evaluate whether the answer can satisfy the question in the matching pattern-based similarity module. 
For the auxiliary task, we employ the $\mathbf{B}^p$ (see Equ. \ref{eq:bert}) from the matching pattern layer for prediction.
We apply a multi-layer perceptron described in Equ. \ref{eq:out} to calculate the similarity score $y^u$ between the user question and the archived answer.
In the same way, we get $y^a$ to represent the archived question answer pair similarity score.
However, due to the lack of question-answer matching labels, we should build the ground-truth for the auxiliary task.
In details,
(1) for each archived question, we regard the corresponding archived answer as the relevant answer; 
(2) for each user question, we regard the corresponding answer with respect to its similar question as the relevant answer.
Thus, we computed the question-answer matching loss $loss^u$ and $loss^a$ with cross-entropy loss again.
The overall loss is defined as the weighted sums of three losses, i.e., 
$$loss=rloss^q+\frac{1-r}{2}loss^u + \frac{1-r}{2}loss^a,$$
where $r \in [0, 1]$ is the main task loss ratio. 
To overcome the issue of sparse irrelevant answers, for each question, we random sample irrelevant answers from its top-$K$\footnote{We set $K=5$ in this paper.} candidate answers which are retrieved from the whole answer collection by BM25 \cite{robertson1995okapi} method.
Note if the answer is irrelevant to both the user question and the archived question, we set $r$ as 0 while training this instance because the answer can not be a bridge in this situation.

In the inference phase, given the user question $Q^u$, the archived question $Q^a$ and the real archived answer $A^a$, we compare the prediction $y^q$ with the threshold \num{0.5} to identify whether the questions are similar or not.

\section{Experiments}

\subsection{Datasets}
We evaluate our model on the following two datasets, i.e., CQADupStack and QuoraQP-a (answer-expanded version of QuoraQP).
The datiled statistics of these datasets are shown in Table \ref{tab:data}.

\begin{table}
    \centering
    \caption{Dataset statistics. \# denotes the number of instances, |$len_Q$| and |$len_A$| denote the average length of the questions and answers, respectively.}
    \begin{tabular}{lccccc}
        \hline
        \hline
         & \#Train & \#Dev & \#Test & |$len_Q$| & |$len_A$| \\
        \hline
        CQADupStack & \num{56633} & \num{5000} & \num{5000} & \num{11.89} & \num{177.70} \\
        QuoraQP-a & \num{281480} & \num{10000} & \num{10000} & \num{13.83} & \num{45.65} \\
        \hline
        \hline
    \end{tabular}
    \label{tab:data}
\end{table}



\begin{itemize}[leftmargin=*]
    \item \textbf{CQADupStack} is a benchmark dataset which is widely used in CQA \cite{hoogeveen2015cqadupstack}. It contains question threads sampled from twelve StackExchange subforums and annotated with similar question information. 
    We take the annotated best answer of the question as the archived answer.
    If there is no best answer for the question, we directly use the answer with the highest score as the archived answer.
    \item \textbf{QuoraQP-a} is built on the widely used CQA dataset QuoraQP\footnote{https://www.kaggle.com/c/quora-question-pairs}, which contains \num{537933} distinct questions from Quora.
    The original dataset cannot be used for our task directly since it does not include archived answers.
    To evaluate our model, we randomly select one question in each pair as the user question and set another one as the archived question.
    Then, we take the top ranked answer from the original website\footnote{https://www.quora.com/} as the archived answer\footnote{We released the dataset at http://tinyurl.com/y8kbbfyu}. 
\end{itemize}

\subsection{Baseline Models}

We compare our proposed model with previous similar question identification methods, which could be classified into two categories based on the usage of answers, i.e., question-only methods and one-side methods.

\subsubsection{\textbf{Question-only Methods}}
Here we consider six existing methods which only rely on questions for similar question identification. 

\begin{itemize}[leftmargin=*]
\item \textbf{TSUBAKI} \cite{shinzato2012tsubaki} accounts for a dependency structure of a sentence and synonyms to evaluate the question similarity.
\item \textbf{BiMPM} \cite{wang2017bilateral} employs a bilateral mechanism to enhance single direction matching in sentence pair relevance modeling.
\item \textbf{ESIM} \cite{chen2016enhanced} is a sequential inference model based on chain LSTMs, which considers the recursive architectures in both local inference modeling and inference composition.
\item  \textbf{DIIN} \cite{gong2017natural} is a instance of Interactive Inference Network (IIN) architecture that hierarchically extracting semantic features from interaction space.
\item \textbf{RE2} \cite{yang2019simple} is a fast and strong neural model with stacked alignment layers, which also employ fusion layer to make the model deeper.
\item \textbf{Bert} \cite{devlin2018bert} is a pre-trained language model based on stacked Transformer \cite{vaswani2017attention} layers, which is effective in measuring the text pair similarity.  
\end{itemize}

\subsubsection{\textbf{One-side Methods}}\label{sec:attn}
We also consider recently proposed methods that employ one-side usage of the archived answer for similar question identification. 

\begin{itemize}[leftmargin=*]

\item \textbf{TSUBAKI+Bert} \cite{sakata2019faq} is a recently proposed method that combine the similarity between questions and the relevance between the user question and archived answer.
\end{itemize}
Here, to fully demonstrate the effectiveness of our model, we also incorporate answers into those question-only methods by two basic operators \cite{gupta2019faq}.
The first one is to directly concatenate the archived question along with its answer. we denote these methods as M$_{concat}$, where M could be any method in the question-only Methods.
The second one is M$_{attn}$, which effectively combines the similarity representations from both the question pair matching and the question answer matching using attention mechanisms in a hierarchical manner.

\subsection{Implementation Details}

We implement our model by Tensorflow \cite{abadi2016tensorflow}.
The hyper-parameters are tuned with the development set.
The model is trained end-to-end by RAdam \cite{liu2019variance} optimizer.
We set the learning rate of RAdam as $5e-5$, and other parameters as $\beta_1=0.9$, $\beta_2=0.999$, $\epsilon=1e-6$. 
We use $3\times3$ kernels in the compression layer.
We use an exponential decayed keep rate during training, where the initial keep rate is \num{1.0} and the decay rate is \num{0.933} for every \num{5000} steps, where the keep rate will achieve to \num{0.5} after \num{50000} steps.
We initial the Bert structure in our model with released Bert-base model\footnote{https://storage.googleapis.com/bert\_models/2018\_10\_18/uncased\_L-12\_H-768\_A-12.zip}.
The other parameters are randomly initialized under a normal distribution with $\mu=0$ and $\sigma=0.2$.
The maximum question length is truncated to 24 for CQADupStack and 32 for QuoraQP-a.
The maximum answer length is truncated to 256 for CQADupStack and 100 for QuoraQP-a.
The batch size is 32 for CQADupStack and 48 for QuoraQP-a. 

For evaluation, we adopt Accuracy, Precision, Recall and F1 score to evaluate the models, and set the Accuracy as the main metric.

\subsection{Hyper-parameter Analysis}

\subsubsection{\textbf{Pattern Similarity Function}}

As described in Section \ref{sec:ps}, we can adopt various pattern similarity functions to calculate the pattern similarity tensor. 
Here, we study the performance of five candidate functions. 
The results are shown in Table \ref{tab:ps}.
As we can see, the choice of pattern similarity function does affect the performance of the Match$^2$ model.
Specifically, the $dot$ function has achieved the best performance in terms of all the evaluation metrics.
The reason might be that the $dot$ function could capture the detailed interactions of each dimension, which are useful in identifying the similarities between matching patterns.
In the following experiments, we will use the $dot$ as the pattern similarity function.

\begin{table}[!t]
  \caption{Results of different similarity functions in the matching pattern-based module on CQADupStack.}
  \label{tab:ps}
  \begin{tabular}{ccccc}
    \hline
    \hline
    Function & Accuracy & Precision & Recall & F1\\
    \hline
    $dot$ & \textbf{62.84} & \textbf{56.34} & \textbf{55.12} & \textbf{55.72} \\
    $cos$ & 62.80 & 56.29 & 55.07 & 55.67 \\
    $l1$ & 62.44 & 55.90 & 54.31 & 55.09 \\
    $l2$ & 62.46 & 55.89 & 54.55 & 55.21 \\
    $jss$ & 62.76 & 56.25 & 54.97 & 55.60 \\
    \hline
    \hline
\end{tabular}
\end{table}

\pgfplotsset{
axis background/.style={fill=white},
grid=both,
  xtick pos=left,
  ytick pos=left,
  tick style={
    major grid style={style=gallery,line width=1pt},
    minor grid style=mercury,
    },
  minor tick num=1,
}

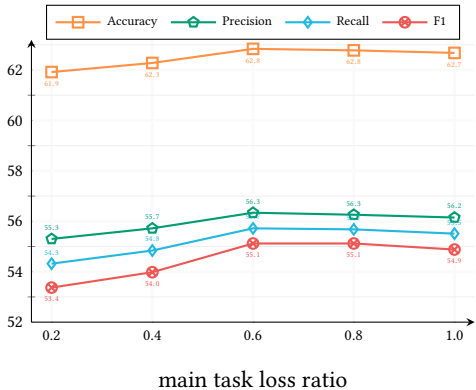
\begin{figure}[!tb]
\centering
  \begin{tikzpicture}
    \begin{axis}[
      height=.3\textwidth,
      width=.42\textwidth,
      xticklabels={0.2,0.4,0.6,0.8,1.0},
      xtick={1,2,3,4,5},
      legend style={
          font=\tiny,
          legend columns=-1,
          at={(0.5,1)},
          anchor=south,
          /tikz/every even column/.append style={column sep=0.9mm}
        },
        ymajorgrids={true},
        ymin=52,
        ymax=63.2,
        ytick={52, 54, ..., 62},
        minor x tick num={0},
        minor y tick num={1},
        axis lines=left,
        enlarge x limits=0.05,
        tickwidth=0pt,
        nodes near coords,
        every node near coord/.append style={anchor=north, font=\fontsize{3pt}{3pt}\selectfont},
        legend entries = {Accuracy, Precision, Recall, F1},
        xlabel={main task loss ratio},
        every tick label/.append style={font=\scriptsize},
        nodes near coords={
        \pgfmathprintnumber[fixed zerofill,precision=1]{\pgfplotspointmeta}
        }
        ]

      \addplot[color=sun_shade,mark=square, every node near coord/.append style={anchor=north}, thick] coordinates {
        (1, 61.92)
        (2, 62.28)
        (3, 62.84)
        (4, 62.78)
        (5, 62.68)
      };
      \addplot[color=free_speech_aquamarine, mark=pentagon, every node near coord/.append style={anchor=south}, thick] coordinates {
        (1, 55.30)
        (2, 55.72)
        (3, 56.34)
        (4, 56.26)
        (5, 56.15)
      };
      \addplot[color=shakespeare, mark=diamond, every node near coord/.append style={anchor=south},thick] coordinates {
        (1, 54.32)
        (2, 54.84)
        (3, 55.72)
        (4, 55.68)
        (5, 55.51)
      };
      \addplot[color=flamingo, mark=otimes, every node near coord/.append style={anchor=north},thick] coordinates {
        (1, 53.37)
        (2, 53.98)
        (3, 55.12)
        (4, 55.12)
        (5, 54.88)
      };
    \end{axis}
  \end{tikzpicture}
  \caption{Results of different main task loss ratios of Match$^2$ on CQADupStack.}
  \label{fig:task}
\end{figure}

\begin{table*}
  \caption{Main Results on CQADupStack and QuoraQP-a. \dag indicates the statistically significant difference over the best baseline model, where +/- indicates the statistically significant improvement/deterioration over the question-only counterpart with $p < 0.01$ \cite{yeh2000more} .}
  \label{tab:res}
  \begin{tabular}{ccl|cccc|cccc}
    \hline
    \hline
     \multicolumn{3}{c}{} & \multicolumn{4}{c}{CQADupStack} & \multicolumn{4}{c}{QuoraQP-a}\\
    Method & \# & Model & Accuracy & Precision & Recall & F1 & Accuracy & Precision & Recall & F1\\
    \hline
     & 1 & TSUBAKI & 56.20 & 50.17 & 34.12 & 40.62 & 66.78 & 51.61 & 36.25 & 42.59 \\
     & 2 & BiMPM & 59.44 & 54.84 & 43.1 & 48.27 & 87.28 & 81.59 & 80.82 & 81.20 \\
    question- & 3 & ESIM & 58.64 & 53.85 & 40.46 & 46.20 & 86.86 & 80.78 & 80.49 & 80.64 \\
     only & 4 & DIIN & 60.30 & 56.13 & 43.83 & 49.22 & 88.01 & 82.48 & 82.20 & 82.33 \\
     & 5 & RE2 & 60.56 & 56.47 & 44.33 & 49.67 & 88.30 & 82.85 & 82.70 & 82.77 \\
     & 6 & Bert & 60.92 & 56.91 & 45.24 & 50.41 & 89.24 & 84.21 & 84.11 & 84.16 \\
    \hline
     & 7 & TSUBAKI+Bert & 57.20$^+$ & 51.75$^+$ & 37.13$^+$ & 43.24$^+$ & 80.23$^+$ & 70.89$^+$ & 70.99$^+$ & 70.94$^+$ \\
     & 8 & BiMPM$_{concat}$ & 59.48 & 54.55 & 46.15$^+$ & 50.00$^+$ & 86.70$^-$ & 80.15$^-$ & 80.91$^+$ & 80.53$^-$ \\
     & 9 & ESIM$_{concat}$ & 59.05$^+$ & 54.35 & 42.14$^+$ & 47.47$^+$ & 86.66$^-$ & 80.09$^-$ & 80.85$^+$ & 80.47$^-$ \\
     & 10 & DIIN$_{concat}$ & 60.74$^+$ & 56.31$^+$ & 47.15$^+$ & 51.33$^+$ & 88.25$^+$ & 82.63 & 82.85$^+$ & 82.74$^+$ \\
     & 11 & RE2$_{concat}$ & 60.16$^-$ & 54.96$^-$ & 51.21$^+$ & 53.02$^+$ & 87.71$^-$ & 79.53$^-$ & 85.91$^+$ & 82.62$^-$ \\
     \textit{one-side} & 12 & Bert$_{concat}$ & 61.50$^+$ & 56.70 & 52.27$^+$ & 54.26$^+$ & 89.81$^+$ & 84.35 & 85.97$^+$ & 85.15$^+$ \\
     & 13 & BiMPM$_{attn}$ & 59.74 & 55.11 & 44.69$^+$ & 49.36$^+$ & 88.18$^+$ & 83.28$^+$ & 81.61$^+$ & 82.44$^+$ \\
     & 14 & ESIM$_{attn}$ & 59.38$^+$ & 54.93$^+$ & 41.59$^+$ & 47.34$^+$ & 87.82$^+$ & 82.76$^+$ & 81.05$^+$ & 81.90$^+$ \\
     & 15 & DIIN$_{attn}$ & 60.78$^+$ & 56.84$^+$ & 44.28 & 49.78 & 88.72$^+$ & 82.76$^+$ & 84.08$^+$ & 83.52$^+$ \\
     & 16 & RE2$_{attn}$ & 61.18$^+$ & 56.55 & 49.93$^+$ & 53.04$^+$ & 89.07$^+$ & 83.32$^+$ & 84.82$^+$ & 84.06$^+$ \\
     & 17 & Bert$_{attn}$ & 61.96$^+$ & 57.31$^+$ & 52.35$^+$ & 54.71$^+$ & 89.92$^+$ & 85.13$^+$ & 85.23$^+$ & 85.18$^+$ \\
    \hline
    \textit{two-side} & 18 & Match$^2$ & \textbf{62.78$^\dag$} & \textbf{58.02$^\dag$} & \textbf{55.03$^\dag$} & \textbf{56.49$^\dag$} & \textbf{90.65$^\dag$} & \textbf{86.21$^\dag$} & \textbf{86.29$^\dag$} & \textbf{86.25$^\dag$} \\
    \hline
    \hline
\end{tabular}
\end{table*}

\subsubsection{\textbf{Multi-task analysis}}

In the model learning phase, we introduced an additional task to train the model.
The final optimization objective of the model is the linear combined loss with pre-defined main task loss ratio $r$.
Here, we study how this ratio affects the model performance. Specifically, we set the weight value from $0.2$ to $1.0$, where the larger value denotes more emphasis on the main task, i.e., similar question identification task.
The results are depicted in Figure~\ref{fig:task}, we can see that there is a consistent tendency between all the evaluation metrics, i.e., the performance first improves along with the increase of the weight value, and drops when the weight become larger than $0.6$. 
The best performance can be obtained at $0.6$, where the model pays balanced attention to both learning objectives.

\begin{table*}
  \caption{Ablation results on CQADupStack and QuoraQP-a. \dag indicates the statistically significant difference over the Match$^2$ model with $p<0.01$ \cite{yeh2000more}.}
  \label{tab:ablation}
  \begin{tabular}{l|cccc|cccc}
    \hline
    \hline
    \multicolumn{1}{l}{} & \multicolumn{4}{c}{CQADupStack} & \multicolumn{4}{c}{QuoraQP-a} \\
    Model & Accuracy & Precision & Recall & F1 & Accuracy & Precision & Recall & F1\\
    \hline
    Match$^2$ & 62.78 & 58.02 & 55.03 & 56.49 & 90.65 & 86.21 & 86.29 & 86.25 \\
    \hline
    Match$^2_Q$ & 60.94$^\dag$ & 56.92$^\dag$ & 45.33$^\dag$ & 50.47$^\dag$ & 89.24$^\dag$ & 84.21$^\dag$ & 84.11$^\dag$ & 84.16$^\dag$ \\
    Match$^2_A$ & 60.44$^\dag$ & 55.46$^\dag$ & 50.25$^\dag$ & 52.72$^\dag$ & 89.11$^\dag$ & 83.32$^\dag$ & 84.97$^\dag$ & 84.14$^\dag$ \\
    Match$^2_{attn}$ & 62.32$^\dag$ & 58.00 & 51.34$^\dag$ & 54.47$^\dag$ & 90.04$^\dag$ & 85.42$^\dag$ & 85.26$^\dag$ & 85.34$^\dag$ \\
    \hline
    \hline
  \end{tabular}
\end{table*}

\begin{figure}[!t]
  \centering
  \includegraphics[scale=0.5]{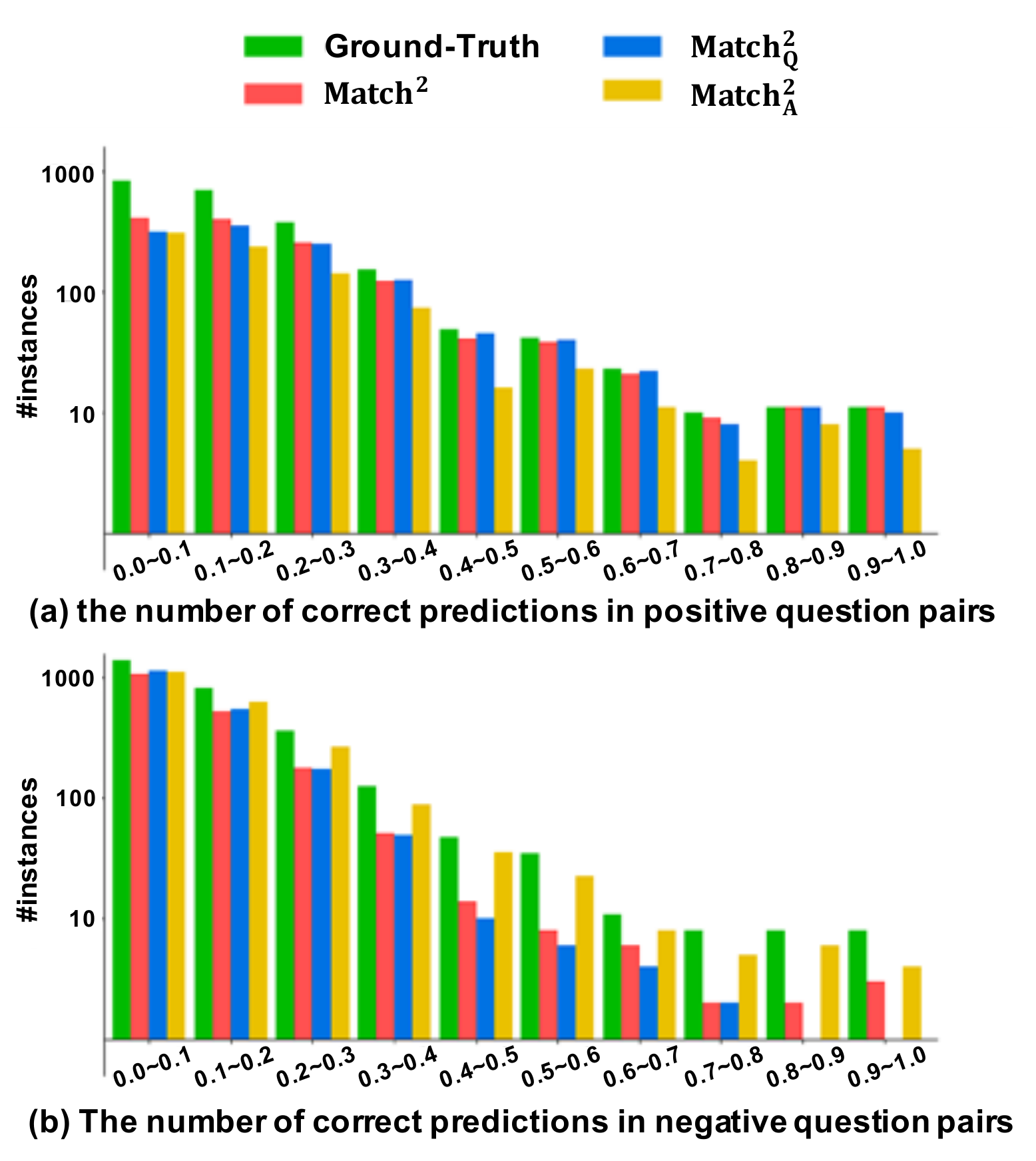}
  \caption{Results on different question groups of CQADupStack. The x-axis represents the Jaccard Index.}
  \label{fig:em}
\end{figure}

\subsection{Main Results}

In this section, we show the main results of the Match$^2$ model as well as baseline methods. All the results are summarized in Table~\ref{tab:res}.

Firstly, for the question-only methods, we can see that neural models (e.g., BiMPM, ESIM and etc.) achieve significant better performance than traditional methods (i.e., TSUBAKI) on both datasets. 
Moreover, it can be observed that the relative improvement of the neural methods over TSUBAKI is much larger on QuoraQP-a than the CQADupStack. 
The reason might that the QuoraQP-a is much larger in size than the CQADupStack, where neural models are often data hungry. 
The Bert achieves the best performance on both datasets in terms of all metrics. 
This indicates the models pre-trained on a large amounts of unstructured texts learn to encode linguistic features that improve the performance.


Secondly, comparing the one-side methods with the question-only methods, we can find that incorporating the answers could indeed improve the performance. 
However, there are also some methods achieving inferior performance with the archived answer. 
For example, the accuracy of RE$_\mathit{concat}$ decreases from $60.56$ to $60.16$ on the CQADupStack. 
This demonstrates that simply incorporating the answers could introduce unexpected noises, which could possibly hurt the performance. 
Moreover, we find that the attention method is relatively more effective than the concatenation method, which indicates the possibility to improve the performance by carefully designed answer usage method.


Thirdly, the Match$^2$ model achieves the best performance in terms of all metrics on both benchmarks. 
For example, the relative improvement of the Match$^2$ model over the best performing baseline method (i.e., Bert$_\mathit{attn}$) is about $3.3\%$ and $1.3\%$ in terms of F1 metric on CQADupStack and QuoraQP-a.
All these demonstrate the effectiveness of the Match$^2$.

\subsection{Analysis on the Match$^2$}\label{sec:ablation}

To better analyze the effect of different components in Match$^2$, we first construct three variants of the model, then evaluate them on both benchmarks and on different question groups.
The constructed variants are listed as follows:

\begin{itemize}[leftmargin=*]

\item \textbf{Match$^2_Q$} is used to represent the representation-based module. It removes the matching pattern-based module and use a multi-layer perceptron (MLP) to replace the aggregation module.

\item \textbf{Match$^2_A$} is for matching pattern-based module. It removes the representation-based module and use a MLP for aggregation.

\item \textbf{Match$^2_{attn}$} adopts the attention mechanism \cite{gupta2019faq} to replace the gate mechanism in the aggregation module,to analyze the effect of the gate mechanism.

\end{itemize}

\subsubsection{\textbf{Sub-Module Analysis}}\label{sec:var}

The performance of different variants are shown in Table~\ref{tab:ablation}. 
Firstly, we can see that both of the Match$^2_Q$ and Match$^2_A$ achieve relatively good performance with the sub-module itself, which demonstrates these modules are effective in most cases. 
Secondly, comparing with these two variants, we find that Match$^2_Q$ achieves higher precision while Match$^2_A$ achieves higher recall. 
This indicates the representation-based module and the matching-pattern based module could be complementary to each other.
Finally, we observe the attention mechanism cannot fully utilize the advantages of the previous modules, which is particularly reflected in the recall metric. 
This difference demonstrates the effectiveness of the gating mechanism in the aggregation component.

\begin{table*}
  \renewcommand{\arraystretch}{1.2}
  \newcommand{\tabincell}[2]{\begin{tabular}{@{}#1@{}}#2\end{tabular}}
    \caption{Two cases from the CQADupStack data. Match$^2_Q$ is the representation-based similarity module, and Match$^2_A$ is the matching pattern-based similarity module.}
    \label{tab:case}
    \begin{tabular}{clcccc}
      \hline
      \hline
       &  & Ground-truth & Match$^2_Q$ & Match$^2_A$ & Match$^2$  \\
      \hline
      Case I & \tabincell{l}{\textbf{$Q^u$}: how to keep a session when logging out \\ \textbf{$Q^a$}: can i close a terminal without killing the command running in it \\ \textbf{$A^a$}: once you log out a terminal, this kill the running session in it as well. \\
      to keep the session alive, you should start a session with `nohup' command. \\
      another way is pause the session with `ctrl-z', pull it into the background \\ with `bg' and then `disown' it.} & 1 & 0 & 1 & 1 \\
      \hline
      Case II & \tabincell{l}{\textbf{$Q^u$}: how do i clean up radiation \\ \textbf{$Q^a$}: how long does radiation take to clean up \\ \textbf{$A^a$}: you have to wait more than 20 years for all the radiation to turn into \\ ground pollution} & 0 & 1 & 0 & 0\\
      \hline
      \hline
  \end{tabular}
  \end{table*}
  
  \begin{figure*}[h]
    \centering
    \includegraphics[scale=0.45]{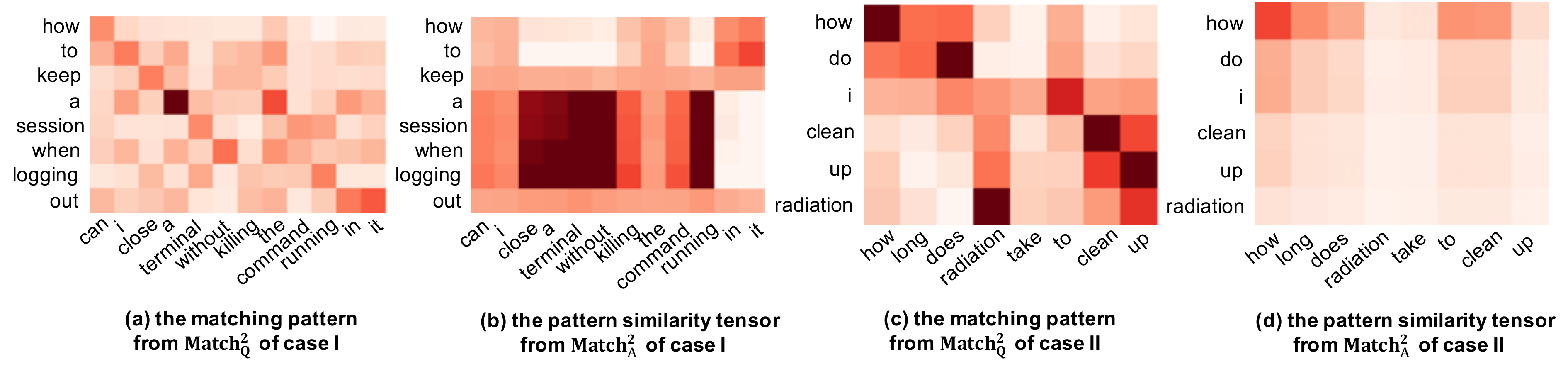}
    \caption{Visualization of the matching pattern and pattern similarity tensor of the above two cases, deeper color indicates higher similarity.}
    \label{fig:vis}
  \end{figure*}

\subsubsection{\textbf{Question Group-based Analysis}}

For more detailed analysis of model performance, we divide the question pairs in CQADupStack into twenty groups based on their similarity and Jaccard Index \cite{jaccard1901etude}, which is a widely-used word level similarity feature. 
We analyze the number of correct predictions in each group. 
The results are shown in Figure~\ref{fig:em}.

We notice that the positive and negative question pairs have similar Jaccard Index distribution in CQADupStack. 
Specifically, for the positive question pairs, we can see the Match$^2_Q$ achieves better performance than the Match$^2_A$ on all the groups, and the gap is larger on the pairs with higher Jaccard Index, i.e. more shared words. 
It indicates that the Match$^2_Q$ could directly capture the word similarities, which is useful to the similar questions with many shared words. 
On the other hand, for the negative question pairs, we can observe that the Match$^2_Q$ could not well address the negative questions pairs with higher Jaccard Index. 
For example, when the Jaccard Index higher than 0.8, the Match$^2_Q$ fails on all the instances. 
The Match$^2_A$ outperforms the Match$^2_Q$ especially on the higher Jaccard Index groups. 
This demonstrates that the matching pattern could avoid the noises from shared words and emphasize the difference between questions. 
Finally, the Match$^2$ module could outperform these two types of modules in most cases. 
It indicates the effectiveness of the gate mechanism that combines the advantage of these two module into a unified model.

\subsection{Case Study and Visualization}

Here, we conduct case studies to better understand what have been learned by the Match$^2$ model.
We also take the Match$^2_Q$ and Match$^2_A$ for comparison. 
The instances are shown in Table~\ref{tab:case}, the first one is a positive question pair with few shared words, while the second one is a negative pair with more common words. 
We can see that the Match$^2_Q$ is not good at dealing with these types of questions, but the Match$^2$ could correctly identify them with the help of Match$^2_A$. 
Specifically, we visualize the matching patterns and pattern similarity tensors from Match$^2_Q$ and Match$^2_A$ in Figure~\ref{fig:vis}. 
For case I, we notice the Match$^2_Q$ is difficult to find out the semantic relation between questions but only recognizes the cluttered similarity presented in Figure \ref{fig:vis}(a). 
By leveraging the archived answer as a bridge, the Match$^2_A$ can easily identify the similarities by comparing the matching patterns, as shown in Figure \ref{fig:vis}(b).

For case II, we notice that the Match$^2_Q$ highlights three similar phrases in Figure \ref{fig:vis}(c), and makes a false positive prediction that the questions are similar.
On the other hand, as shown in Figure \ref{fig:vis}(d), the only semantic relevance between two questions is ``how'' which means the questions are different except their question type.
Based on the pattern similarity tensor, the Match$^2_A$ is able to predict these two questions as different.

\section{Conclusion and Future Work}
In this paper, we introduced a two-side usage of the archived answer for similar question identification task by leveraging the answer as a bridge of the questions.
We proposed a novel matching over matching (Match$^2$) model, which consists of three main components, namely the \textit{representation-based similarity module}, \textit{matching pattern-based similarity module}, and the \textit{aggregation module}.
Empirical experiments on two benchmarks demonstrate that our model can significantly outperform previous state-of-the-art methods. Moreover, we also conducted rigorous experiments on the sub-modules to verify the effectiveness of the model.
In the future work, we would like to extend our model to leverage variant number of answers and take the answer quality into account.

\begin{acks}
This work was supported by the National Natural Science Foundation of China (NSFC) under Grants No. 61722211, 61773362, 61872338, and 61902381, Beijing Academy of Artificial Intelligence (BAAI) under Grants No. BAAI2019ZD0306, and BAAI2020ZJ0303, the Youth Innovation Promotion Association CAS under Grants No. 20144310, and 2016102, the National Key RD Program of China under Grants No. 2016QY02D0405, the Lenovo-CAS Joint Lab Youth Scientist Project, and the Foundation and Frontier Research Key Program of Chongqing Science and Technology Commission (No. cstc2017jcyjBX0059).
\end{acks}

\bibliographystyle{ACM-Reference-Format}
\bibliography{ref}

\end{document}